\documentclass{article}
\usepackage[dvips]{graphicx}
\topmargin=-1cm\oddsidemargin=0truecm\evensidemargin=0truecm\textheight=24.0cm\textwidth=16.5cm\usepackage{color}
\def\vol#1#2#3{{\bf {#1}} ({#2}) {#3}}\def\NP{Nucl.~Phys. }\def\PL{Phys.~Lett. }\def\PR{Phys.~Rev. }\def\PRP{Phys.~Rep. }\def\PRL{Phys.~Rev.~Lett. }\def\PTP{Prog.~Theor.~Phys. }\def\IJMP{Int.~J.~Mod.~Phys. }\def\JP{J.~Phys. }

\def\no{\nonumber}

\def\2tvec#1#2{
\left(
\begin{array}{c}
#1  \\
#2  \\   
\end{array}
\right)}

\def\mat2#1#2#3#4{
\left(
\begin{array}{cc}
#1 & #2 \\
#3 & #4 \\
\end{array}
\right)
}

\def\Mat3#1#2#3#4#5#6#7#8#9{
\left(
\begin{array}{ccc}
#1 & #2 & #3 \\
#4 & #5 & #6 \\
#7 & #8 & #9 \\
\end{array}
\right)
}

\def\3tvec#1#2#3{
\left(
\begin{array}{c}
#1  \\
#2  \\   
#3  \\
\end{array}
\right)}

\def\4tvec#1#2#3#4{
\left(
\begin{array}{c}
#1  \\
#2  \\   
#3  \\
#4  \\
\end{array}
\right)}

\def\L{\left}
\def\R{\right}

\def\hbar{\hspace{1mm}\bar{}\hspace{-1mm}h}

\def\eqn#1{
\begin{eqnarray}
#1
\end{eqnarray}
}

\begin{document}

\begin{titlepage}
\begin{flushright}
KIAS-P12017
\end{flushright}
\begin{center}

\vspace{1cm}
{\Large\bf Suppressing Proton Decay by Cancellation
 in $S_4$ Flavor Symmetric Extra U(1) Model}\\
 \vspace{0.5cm}
 Yasuhiro Daikoku$^{a}$\footnote{yasu\_daikoku@yahoo.co.jp} and
Hiroshi Okada$^{b}$\footnote{hokada@kias.re.kr}
\vspace{5mm}

 {\it 
$^a${ Institute for Theoretical Physics, Kanazawa University, Kanazawa 920-1192, Japan} \\ 
 \vspace{1mm} 
 $^b${School of Physics, KIAS, Seoul 130-722, Korea}\\ \vspace{1mm}
}
%%\maketitle
  
  \vspace{8mm}

\begin{abstract}
We consider proton stability based on 
$E_6$ inspired extra U(1) model with $S_4$ flavor symmetry.
In this model, a long life time of proton is realized by the flavor symmetry.
One of the interesting effects of flavor symmetry is that
the proton decay widths of $p\to\mu^+ X$ are suppressed by cancellation. 
This suppression mechanism is important in the case that
Yukawa coupling constants are hierarchical.
Our model predicts $p\to e^+K^0$ has larger  decay width  than that of
$p\to\mu^+K^0$.
 
\end{abstract}
%%%%

\end{center}
\end{titlepage}
\setcounter{footnote}{0}

\newpage

\section{Introduction}

Supersymmetry is an elegant solution of hierarchy problem of Standard Model (SM) \cite{SUSY},
however, a simple supersymmetric extension of SM suffers  from non-conservation of baryon number
and $\mu$-problem. 
Therefore we must introduce new symmetry such as R-parity in MSSM,
to suppress proton decay operators and $\mu$-term.
One of the solutions of $\mu$-problem is given by introducing extra $U(1)$ gauge symmetry 
\cite{extra-u1}. 
In this frame work, several new superfields such  as singlet $S$, exotic quarks $G,G^c$,
must be introduced to cancel gauge anomaly.
This is the elegant solution of $\mu$-problem, 
however proton instability is not solved because
the baryon number violating interactions in superpotential of MSSM are replaced by  single
exotic quark interactions.

In the superpotential,  there is no obvious distinction between
baryon number violating trilinear terms and Yukawa  interactions.
Therefore it is natural to introduce flavor symmetry in suppressing baryon number
violating operators. 
Especially non-Abelian discrete symmetry is good candidate for the flavor symmetry,
because large mixing angles of Maki-Nakagawa-Sakata (MNS) mixing matrix
may be explained simultaneously, and more simply, non-Abelian symmetry can be the reason why generation exists.
At previous work,
we explained $S_4\times Z_2$ flavor symmetry not only realizes maximal mixing angle $\theta_{23}$
but also suppresses proton decay based on
$SU(3)_c\times SU(2)_W\times U(1)_Y\times U(1)_X\times U(1)_Z$ gauge symmetry \cite{s4u1}.
As the suppression mechanism of proton decay is complicated and model dependent,
in this paper, we give more detailed estimation of proton life time and 
investigate several flavon sectors.

%In the several models of the non-Abelian discrete symmetries,
%although the direction of vacuum expectation values (VEVs) of flavons is crucial,
%the direction is selected by hand adding the explicit symmetry breaking terms.
%As such procedure is unnatural, 

If we start from exactly flavor symmetric theory,
the spontaneous flavor symmetry breaking realizes special VEV direction of flavons, 
which affects proton life time significantly.
If the Yukawa hierarchy is realized by Froggatt-Nielsen mechanism,
as $p\to e^+X$ are suppressed 
due to the small coupling  constants, $p\to \mu^+X$ may dominate proton decay width.
In our model, as $p\to \mu^+X$ are suppressed by cancellation,
$p\to e^+K^0$ dominates the proton decay width.

This paper is organized  as follows.
In section 2, we estimate proton life time based on gauge non-singlet flavon model.
In section 3, we modify the flavon sector by adding gauge-singlet flavon and
Froggatt-Nielsen flavon.
In section 4, we eliminate gauge non-singlet flavon and construct
Dirac neutrino model. 
Finally we give conclusion of our analysis in section 5.

\section{$S_4\times Z_2$ flavor symmetric extra U(1) model}

At first we explain the basic structure of our model.
We extend the gauge symmetry to $G_{32111}=SU(3)_c\times SU(2)_W\times U(1)_Y\times U(1)_X\times U(1)_Z$
which is the subgroup of $E_6$.
In order to cancel gauge anomaly, we must add new superfields, such as
SM singlet $S$, exotic quark $G,G^c$ (hereafter we call them g-quark) and right handed neutrino (RHN) $N^c$.
We can embed these superfields with MSSM superfields $Q,U^c,D^c,L,E^c,H^U,H^D$
into ${\bf 27}$ of $E_6$ \cite{e6}.
As the singlet  $S$ develops VEV and breaks $U(1)_X$ gauge symmetry, $O$ (TeV) scale $\mu$-term
is induced naturally.
In order to break $U(1)_Z$ and generate a large Majorana mass of RHN,
we add SM singlet $\Phi,\Phi^c$.
The gauge representations of superfields are given in Table 1 \cite{s4u1}.

\begin{table}[htbp]
\begin{center}
\begin{tabular}{|c|c|c|c|c|c|c|c|c|c|c|c||c|c|}
\hline
         &$Q$ &$U^c$    &$E^c$&$D^c$    &$L$ &$N^c$&$H^D$&$G^c$    
&$H^U$&$G$ &$S$ &$\Phi$&$\Phi^c$\\ \hline
$SU(3)_c$&$3$ &$3^*$    &$1$  &$3^*$    &$1$ &$1$  &$1$  &$3^*$    &$1$  
&$3$ &$1$ &$1$   &$1$     \\ \hline
$SU(2)_W$&$2$ &$1$      &$1$  &$1$      &$2$ &$1$  &$2$  &$1$      &$2$  
&$1$ &$1$ &$1$   &$1$     \\ \hline
$y=6Y$   &$1$ &$-4$     &$6$  &$2$      &$-3$&$0$  &$-3$ &$2$      &$3$  
&$-2$&$0$ &$0$   &$0$     \\ \hline
$x$      &$1$ &$1$      &$1$  &$2$      &$2$ &$0$  &$-3$ &$-3$     &$-2$ 
&$-2$&$5$ &$0$   &$0$     \\ \hline
$z$      &$-1$&$-1$     &$-1$ &$2$      &$2$ &$-4$ &$-1$ &$-1$     &$2$  
&$2$ &$-1$&$8$   &$-8$    \\ \hline
$R$      &$-$ &$-$      &$-$  &$-$      &$-$ &$-$  &$+$  &$+$      &$+$  
&$+$ &$+$ &$+$   &$+$     \\ \hline
\end{tabular}
\end{center}
\caption{$G_{32111}$ assignment of fields.
Where the $x$, $y$ and $z$ are charges of $U(1)_X$, $U(1)_Y$ and $U(1)_Z$, 
and $Y$ is hypercharge. R-parity $R=\exp\L[\frac{i\pi}{20}(3x-8y+15z)\R]$ is unbroken.}
\end{table}

Under the gauge symmetry given in Table 1, the renormalizable superpotential is given by
\eqn{
W&=& Y^UH^UQU^c +Y^DQD^cH^D +Y^EH^DLE^c +Y^NH^ULN^c +Y^M\Phi N^cN^c+\lambda SH^UH^D+kSGG^c \no \\
&+&M\Phi \Phi^c +Y^{QQG}GQQ +Y^{UDG}G^cU^cD^c +Y^{EUG}GE^cU^c +Y^{QLG}G^cLQ +Y^{NDG}GN^cD^c.
}
In this superpotential, unwanted terms are included in the second line.
The first term of the second line is the mass term of singlets $\Phi, \Phi^c$ which prevent singlets
from developing VEVs. The other five terms of the second line are single g-quark interactions, which
break baryon and lepton number and induce rapid proton decay.
In the first line, we must take care of the flavor changing neutral currents (FCNCs)
induced by extra Higgs bosons \cite{e6-FCNC}.
Therefore the superpotential Eq.(1) is not consistent at the present stage.

In order to stabilize proton, we introduce $S_4\times Z_2$ flavor symmetry.
If we assign $G,G^c$ to $S_4$ triplet and quarks and leptons to doublet or singlet,
the single g-quark interaction is forbidden. 
However, as the g-quark must never be stable from phenomenological reason, 
we assign $\Phi^c$ to $S_4$ triplet to break the flavor symmetry slightly.
In this case, as $\Phi,\Phi^c$ play the role of flavons, we call them gauge non-singlet flavons. 
In order to realize the large mixing angle of $\theta_{23}$ in the MNS matrix and
suppress the Higgs-mediated FCNCs, we assign the superfields 
in our model as given in Table 2 \cite{s4pamela}.

In the non-renormalizable part of superpotential, 
the single g-quark interactions which contribute to the g-quark decay are given as follows 
\eqn{
W&\supset&\frac{Y^{QQG}}{M^2_P}\Phi\Phi^cQQG
+\frac{Y^{UDG}}{M^2_P}\Phi\Phi^cG^cU^cD^c
+\frac{Y^{EUG}}{M^2_P}\Phi\Phi^cGE^cU^c
+\frac{Y^{QLG}}{M^2_P}\Phi\Phi^cG^cLQ .
}

\begin{table}[htbp]
\begin{center}
\begin{tabular}{|c|c|c|c|c|c|c|c|c|c|}
\hline
        &$Q_1$    &$Q_2$    &$Q_3$     &$U^c_1$  &$U^c_2$  &$U^c_3$   &$D^c_1$  &$D^c_2$  &$D^c_3$ \\
      \hline
$S_4$   &${\bf 1}$&${\bf 1}$& ${\bf 1}$&${\bf 1}$&${\bf 1}$&${\bf 1}$ &${\bf 1}$&${\bf 1}$&${\bf 1}$\\
      \hline
$Z_2$   &$-$      &$-$      &$-$       &$-$      &$-$      &$-$       &$-$      &$-$      &$-$ \\
      \hline
      \hline
        &$E^c_1$  &$E^c_2$  &$E^c_3$   &$L_i$    &$L_3$    &$N^c_i$   &$N^c_3$  &$H^D_i$  &$H^D_3$  \\
      \hline
$S_4$   &${\bf 1}$&${\bf 1}$&${\bf 1'}$&${\bf 2}$&${\bf 1}$&${\bf 2}$ &${\bf 1}$&${\bf 2}$&${\bf 1}$ \\
      \hline      
$Z_2$   &$+$      &$-$      &$+$       &$-$      &$-$      &$+$       &$-$      &$-$      &$+$ \\
      \hline
      \hline   
        &$H^U_i$  &$H^U_3$  &$S_i$     &$S_3$    &$G_a$    &$G^c_a$   &$\Phi_i$ &$\Phi_3$ &$\Phi^c_a$\\
      \hline
$S_4$   &${\bf 2}$&${\bf 1}$&${\bf 2}$ &${\bf 1}$&${\bf 3}$&${\bf 3}$ &${\bf 2}$&${\bf 1}$&${\bf 3}$\\
      \hline
$Z_2$   &$-$      &$+$      &$-$       &$+$      &$+$      &$+$       &$+$      &$+$      &$+$ \\
      \hline
\end{tabular}
\end{center}
\caption{$S_4\times Z_2$ assignment of superfields
(Where the index $i$ of the $S_4$ doublets runs $i=1,2$,
and the index $a$ of the $S_4$ triplets runs $a=1,2,3$.)}
\end{table}

\subsection{Higgs sector and hidden sector}

Under the flavor symmetry given in Table 2, the superpotential of Higgs sector is given by,
\eqn{
W_H&=&\lambda_1S_3(H^U_1H^D_1+H^U_2H^D_2)+\lambda_3S_3H^U_3H^D_3  \no \\
&+&\lambda_4H^U_3(S_1H^D_1+S_2H^D_2)+\lambda_5(S_1H^U_1+S_2H^U_2)H^D_3.
}
where one can take, without any loss of the generalities, 
$\lambda_{1,3,4,5}$ as real, by redefining four arbitrary fields of 
$\{S_i, S_3, H^U_i, H^U_3, H^D_i, H^D_3\}$.
As only $S_4$ singlets $H^U_3,H^D_3$ and $S_3$ couple to quarks and g-quarks respectively, 
they behave like MSSM Higgs and SM singlet respectively. $S_i$ also behaves like SM singlets.
Through the
renormalization group equations, the squared masses of $H^U_3,S_3$ become negative,
they develop VEVs and break gauge symmetry. 
As the result, the A-term $A_3S_3H^U_3H^D_3$ enforces  $H^D_3$ developing VEV.
However, $S_4$ doublets do not develop VEVs.
To generate the VEVs of $S_4$ doublets, we must add flavor breaking squared mass terms.
The origin of flavor breaking terms is discussed below. 
Note that there is accidental $O(2)$ symmetry 
induced by the common rotation of the $S_4$ doublets in $W_H$.

As it is thought that the scalar squared masses are induced 
as the result of SUSY breaking in hidden sector,
we assume flavor symmetry is broken at the same time.
We assume hidden sector is described by
flavor symmetric extension of O'Raifeartaigh model \cite{OR}.
We introduce gauge singlet $A,B_+,B_i,C_+,C_i$ and assign 
$Z'_2$ charges to them to separate hidden sector from observable sector.
We assume $U(1)_R$ symmetry is hold at the limit of infinite Planck scale,
$M_P\to \infty$. The representations of hidden sector superfields
under the $S_4\times Z_2\times Z'_2\times U(1)_R$ symmetry are given in table 3.

\begin{table}[htbp]
\begin{center}
\begin{tabular}{|c|c|c|c|c|c|}
\hline
          &$A$    &$B_+$   &$B_i$     &$C_+$   &$C_i$ \\  
\hline        
$S_4$     &$1$    &$1$     &$2$       &$1$     &$2$   \\ 
\hline
$Z_{2(4)}$&$+(0)$ &$+(0)$  &$-(1/2)$  &$+(0)$  &$-(1/2)$   \\ 
\hline
$Z'_2$    &$+$    &$-$     &$-$       &$-$     &$-$   \\ 
\hline
$U(1)_R$  &$2$    &$2$     &$2$       &$0$     &$0$   \\
\hline
\end{tabular}
\end{center}
\caption{$S_4\times Z_{2(4)}\times Z'_2\times U(1)_R$ assignment of superfields
(Where the index $i$ of the $S_4$ doublets runs $i=1,2$.)}
\end{table}

Under the symmetry given in Table 3, the superpotential of hidden sector is given by,
\eqn{
W_{\mbox{hidden}}&=&-M^2A+m_+B_+C_++m(B_1C_1+B_2C_2)+\frac12\lambda_+AC^2_+
+\frac12\lambda A(C^2_1+C^2_2) .
}
For the F-terms of hidden sector superfields,
\eqn{
F_A&=&-M^2+\frac12\lambda_+C^2_+ +\frac12\lambda(C^2_1+C^2_2), \\
F_{B_+}&=&m_+C_+ ,\\
F_{B_1}&=&mC_1 ,\\
F_{B_2}&=&mC_2 ,\\
F_{C_+}&=&m_+B_++\lambda_+AC_+ ,\\
F_{C_1}&=&mB_1+\lambda AC_1 ,\\
F_{C_2}&=&mB_2+\lambda AC_2, 
}
as it is impossible to satisfy the equations $F_A=0$ and $F_{B_+}=F_{B_i}=0$ at the same time,
SUSY is spontaneously broken.
At the same time, flavor symmetry $S_4\times Z_2$ is also broken spontaneously.
As the superpotential has accidental $O(2)$ symmetry,
the direction of $S_4$ doublet F-term defined by,
\eqn{
F_{B_1}=Fc_B,\quad F_{B_2}=Fs_B,
}
is described by free parameter $\theta_B$. Although  the spontaneous breaking of $O(2)$ results the
appearance  of Nambu-Goldstone boson (NGB), as the interactions of  the NGB 
with observable  sector particles are suppressed by Planck scale and the mixings between Higgs bosons and NGB are suppressed by large hidden sector VEV scale, 
we assume this NGB does not 
cause any problem. The problem of R-axion, which is the NGB of spontaneous $U(1)_R$ symmetry breaking,
may be avoided by adding explicit $U(1)_R$ symmetry breaking higher dimensional terms \cite{R-sym}.
The problem of O(2) NGB may be also solved by adding O(2) breaking
higher dimendional terms.

The SUSY breaking in hidden sector is mediated to observable sector by gravity
through the non-renormalizable terms in K\"ahler potential as given by,
\eqn{
K\supset \frac{1}{M^2_P}[a_H(H_1B_1+H_2B_2)(H_1B_1+H_2B_2)^\dagger
+(b_HB_+H_3(H_1B_1+H_2B_2)^\dagger+h.c. )],
}
where $H=H^U,H^D,S$, and flavor symmetric terms and the contributions to other superfields are omitted
\footnote{In this paper, we assume flavor symmetric SUSY breaking parameters are larger than 
flavor breaking SUSY breaking parameters without any reason, because we are interested in the effects of degenerated scalar g-quark mass spectrum.}.
These terms induce scalar squared mass terms as follow,
\eqn{
V_{FB}= m^2_{BH1}|c_BH_1+s_BH_2|^2
+[m^2_{BH2}H_3(c_BH_1+s_BH_2)^\dagger+h.c.].
}
If we substitute the VEV $\L<H_1\R>=vc_H, \L<H_2\R>=vs_H,\L<H_3\R>=v'$ for $H_a$, then we get
\eqn{
V_{FB}&=&m^2_{BH1}v^2\cos^2(\theta_B-\theta_H)
+[m^2_{BH2}vv'\cos(\theta_B-\theta_H)+h.c] \no \\
&=&a[\cos(\theta_B-\theta_H)-b]^2+\mbox{const} .
}
At second line, we simplified the equation, because we are interested only in direction $\theta_H$.
As we can change the sign of $b$ by the redefinition of the sign of $v$,
we can define $b>0$ without loss of generality.
The minimum of potential $V_{FB}$ is classified as follow,
\eqn{
a>0, b>1&:&\theta_H=\theta_B, \\
a>0 ,b<1&:&\theta_H=\theta_B-\arccos b, \\
a<0&:&\theta_H=\theta_B+\pi ,
}
from which  one can see that the angle $\theta_H$ is controlled by 
free parameter $a,b$ for the given $\theta_B$.
In this section, we assume the condition Eq.(16) is satisfied for $H^U,H^D,S$ and
select the common  VEV direction $\theta_B$.
In this direction, as the flavor symmetric part of Higgs potential has accidental $O(2)$ symmetry
and depends only on $\theta_{H^U}-\theta_{H^D},\theta_{H^U}-\theta_S,\theta_{H^D}-\theta_S$ ,
$\theta_H(=\theta_{H^U}=\theta_{H^D}=\theta_S=\theta_B)$ is fixed by $V_{FB}$.
Therefore we can not assume $V_{FB}$ as perturbation
even if flavor breaking parameter is small.

\subsection{Flavon sector}

The superpotential of flavon sector is given by,
\eqn{
W_\Phi&=&\frac{Y^\Phi_1}{2M_P}\Phi^2_3\L[(\Phi^c_1)^2+(\Phi^c_2)^2+(\Phi^c_3)^2\R] \no \\
&+&\frac{Y^\Phi_2}{2M_P}(\Phi^2_1+\Phi^2_2)\L[(\Phi^c_1)^2+(\Phi^c_2)^2+(\Phi^c_3)^2\R] \no \\
&+&\frac{Y^\Phi_3}{2M_P}\L\{2\sqrt{3}\Phi_1\Phi_2\L[(\Phi^c_2)^2-(\Phi^c_3)^2\R]
+(\Phi^2_1-\Phi^2_2)\L[(\Phi^c_2)^2+(\Phi^c_3)^2-2(\Phi^c_1)^2\R]\R\} \no \\
&+&\frac{Y^\Phi_4}{2M_P}\Phi_3\L\{\sqrt{3}\Phi_1\L[(\Phi^c_2)^2-(\Phi^c_3)^2\R]
+\Phi_2\L[(\Phi^c_2)^2+(\Phi^c_3)^2-2(\Phi^c_1)^2\R]\R\},
}
and the flavor symmetric part of potential is given by,
\eqn{
V&=&-m^2_1|\Phi_3|^2+m^2_2[|\Phi_1|^2+|\Phi_2|^2]
+m^2_3[|\Phi^c_1|^2+|\Phi^c_2|^2+|\Phi^c_3|^2] \no \\
&-&\frac{A^\Phi_1}{2M_P}\Phi^2_3\L[(\Phi^c_1)^2+(\Phi^c_2)^2+(\Phi^c_3)^2\R] \no \\
&-&\frac{A^\Phi_2}{2M_P}(\Phi^2_1+\Phi^2_2)\L[(\Phi^c_1)^2+(\Phi^c_2)^2+(\Phi^c_3)^2\R] \no \\
&-&\frac{A^\Phi_3}{2M_P}\L\{2\sqrt{3}\Phi_1\Phi_2\L[(\Phi^c_2)^2-(\Phi^c_3)^2\R]
+(\Phi^2_1-\Phi^2_2)\L[(\Phi^c_2)^2+(\Phi^c_3)^2-2(\Phi^c_1)^2\R]\R\} \no \\
&-&\frac{A^\Phi_4}{2M_P}\Phi_3\L\{\sqrt{3}\Phi_1\L[(\Phi^c_2)^2-(\Phi^c_3)^2\R]
+\Phi_2\L[(\Phi^c_2)^2+(\Phi^c_3)^2-2(\Phi^c_1)^2\R]\R\} \no \\
&+&\mbox{F-term}+\mbox{D-term}.
}
As this potential does not have accidental continuum symmetry, 
flavor breaking terms can be assumed as perturbation.
We assume  negative squared mass of $\Phi_3$ pulls the trigger of $U(1)_Z$ breaking
and $\Phi_3$ and $\Phi^c_a$ develop VEVs along  the D-flat direction.
For the VEV of $\Phi_i$, there are two possibilities  in the flavor symmetry breaking.
The one of them is $S_3$ symmetric vacuum defined as,
\eqn{
\Phi_3=V,\quad \Phi_i=0,\quad \Phi^c_a=\frac{V}{\sqrt{3}}(1,1,1),
}
and the other is $S_3$ breaking vacuum defined as,
\eqn{
\Phi_3=Vc,\quad \Phi_i=Vs(0,1),\quad \Phi^c_a=\frac{V}{\sqrt{2}}(0,1,1).
}
 Other degenerated vacuums are given by acting $S_4$ translations on these VEVs, respectively.
Note that spontaneous breaking of discrete symmetry causes domain wall problem.
We assume flavor symmetry is not recovered  in reheating era after inflation \cite{domain-wall}
\footnote{Here we note that the reheating temerature must be lower  than $10^7$ GeV
to avoid over production of gravitino in gravity mediation scenario \cite{g-lifetime}.
For such a low temperature, there is no particle such as $U(1)_Z$ gauge boson or RHN
which interacts with flavons through renormalizable operators
in thermal bath. Therefore thermal mass of flavon is induced by nonrenormalizable terms
and negligible  (for example, $m_{\mbox{thermal}} \sim m_\nu T/V<10^{-6}$ eV for the effective operator $\frac{m_\nu}{V}\Phi\nu\nu$) , and flavor symmetry is not recovered.}. 
As the vacuum defined in Eq.(22) is not phenomenologically allowed 
because g-quark $g_1,g^c_1$ become stable,
we select the vacuum  defined in Eq.(21) by tuning  $m^2_2$ large  enough.

From the simplified potential
\eqn{
V\sim -m^2_{SUSY}|\Phi|^2+\frac{|Y^\Phi|^2}{M^2_P}|\Phi^3|^2,
}
the size of flavon VEV is estimated as follow,
\eqn{
\Phi \sim \sqrt{\frac{m_{SUSY}M_P}{Y^\Phi}}.
}
If we put $Y^\Phi\sim 0.01$ and $m_{SUSY}\sim 10$ TeV, then we get $V\sim 10^{12}$ GeV
which is the favorable value to satisfy the constraint for g-quark and proton life time 
at the same time \cite{king}.

\subsection{Quark and Lepton sector}

As the flavor symmetry reduces the number of free parameters drastically,
we can decide the value of free parameters by very few assumptions.
For the quark sector, superpotential is given by
\eqn{
W_Q=Y^U_{ab}H^U_3Q_aU^c_b+Y^D_{ab}H^D_3Q_aD^c_b.
}
As the extra Higgs $H^U_i,H^D_i$ do not couple to quarks,
Higgs mediated flavor changing neutral currents are not induced.

For the lepton sector, superpotential is given by
\eqn{
W_L&=&Y^N_2\L[H^U_1(L_1N^c_2+L_2N^c_1)+H^U_2(L_1N^c_1-L_2N^c_2)\R] \no \\
&+&Y^N_3H^U_3L_3N^c_3+Y^N_4L_3(H^U_1N^c_1+H^U_2N^c_2) \no \\
&+&Y^E_1E^c_1(H^D_1L_1+H^D_2L_2)+Y^E_2E^c_2H^D_3L_3+Y^E_3E^c_3(H^D_1L_2-H^D_2L_1) \no \\
&+&\frac12 Y^M_1\Phi_3(N^c_1N^c_1+N^c_2N^c_2)+ \frac12 Y^M_3\Phi_3 N^c_3N^c_3 \no \\
&+&\frac12Y^M_2[\Phi_1(2N^c_1N^c_2)+\Phi_2(N^c_1N^c_1-N^c_2N^c_2)].
}
Without any loss of generalities,  by the field redefinition, we can define
$Y^E_{1,2,3},Y^M_{1,3},Y^N_{2,4}$ are real and non-negative and
$Y^M_2,Y^N_3$ are complex.
We tune the angle $\theta_B=\theta_{23}=\frac{\pi}{4}$ by hand and
define the VEVs of scalar fields as follows
\eqn{
&&\L<H^U_1\R>=\L<H^U_2\R>=\frac{1}{\sqrt{2}}v_u,\quad 
\L<H^U_3\R>=v'_u ,\quad 
\L<H^D_1\R>=\L<H^D_2\R>=\frac{1}{\sqrt{2}}v_d,\quad 
\L<H^D_3\R>=v'_d ,\no \\
&&\L<S_1\R>=\L<S_2\R>=\frac{1}{\sqrt{2}}v_s,\quad
\L<S_3\R>=v'_s ,\no \\
&&\L<\Phi_1\R>=\L<\Phi_2\R>=0,\quad
\L<\Phi_3\R>=V , \quad
\L<\Phi^c_1\R>=\L<\Phi^c_2\R>=\L<\Phi^c_3\R>=\frac{V}{\sqrt{3}},
}
and define the mass parameters as follows \cite{kubo}
\eqn{
\begin{tabular}{llll}
$M_1=Y^M_1V$,        & $M_3=Y^M_3V$,          &                     &   \\
$m^\nu_2=Y^N_2v_u$,  & $m^\nu_3=|Y^N_3|v'_u$, & $m^\nu_4=Y^N_4v_u$, &   \\
$m^l_1=Y^E_1v_d$,    & $m^l_2=Y^E_2v'_d$,     & $m^l_3=Y^E_3v_d$.   &
\end{tabular}
}
With these parameters, the mass matrices are given by
\eqn{
\begin{tabular}{ll}
$M_l=\frac{1}{\sqrt{2}}\Mat3{m^l_1}{0}{-m^l_3}{m^l_1}{0}{m^l_3}{0}{\sqrt{2}m^l_2}{0}$, &
$M_D=\frac{1}{\sqrt{2}}\Mat3{m^\nu_2}{m^\nu_2}{0}{m^\nu_2}{-m^\nu_2}{0}
{m^\nu_4}{m^\nu_4}{\sqrt{2}e^{i\delta}m^\nu_3}$,  \\
$M_R=\Mat3{M_1}{0}{0}{0}{M_1}{0}{0}{0}{M_3}$. &
\end{tabular}
}
Due to the seesaw mechanism, the neutrino mass matrix is given by
\eqn{
M_\nu&=&M_DM^{-1}_RM^t_D=\Mat3{\rho^2_2}{0}{\rho_2\rho_4}
{0}{\rho^2_2}{0}
{\rho_2\rho_4}{0}{\rho^2_4+e^{2i\delta}\rho^2_3},
}
where
\eqn{
\rho_2=\frac{m^\nu_2}{\sqrt{M_1}},\quad \rho_4=\frac{m^\nu_4}{\sqrt{M_1}},\quad \rho_3=\frac{m^\nu_3}{\sqrt{M_3}}.
}
The charged lepton mass matrix is diagonalized as follow
\eqn{
V^\dagger_l M^*_l M^t_l V_l&=&\mbox{diag}(m^2_e,m^2_\mu, m^2_\tau)=((m^l_2)^2,(m^l_3)^2,(m^l_1)^2), \\
V_l&=&\frac{1}{\sqrt{2}}\Mat3{0}{-1}{1}
{0}{1}{1}
{-\sqrt{2}}{0}{0},
}
and those of the light neutrinos are given by
\eqn{
V^t_\nu M_\nu V_\nu&=&\mbox{diag}(e^{i(\phi_1-\phi)}m_{\nu_1},e^{i(\phi_2+\phi)}m_{\nu_2},m_{\nu_3}), \\
V_\nu&=&
\Mat3{-\sin\theta_\nu}{e^{i\phi}\cos\theta_\nu}{0}
{0}{0}{1}
{e^{-i\phi}\cos\theta_\nu}{\sin\theta_\nu}{0}.
}
From the above equations, the MNS matrix is given by
\eqn{
V_{MNS}&=&V^\dagger_lV_\nu P_\nu
=\frac{1}{\sqrt{2}}\Mat3{-\sqrt{2}e^{-i\phi}\cos\theta_\nu}{-\sqrt{2}\sin\theta_\nu}{0}
{\sin\theta_\nu}{-e^{i\phi}\cos\theta_\nu}{1}
{-\sin\theta_\nu}{e^{i\phi}\cos\theta_\nu}{1}P_\nu,
}
where
\eqn{
P_\nu=\mbox{diag}(e^{-i(\phi_1-\phi)/2},e^{-i(\phi_2+\phi)/2},1).
}
%%% 
Here it is worth mentioning that a rather large mixing angle of 1-3 component of $V_{MNS}$; $\theta_{13}$, is measured by
the recent experiments such as T2K \cite{t2k11}, Double Chooz \cite{dc11}, Daya-Bay \cite{daya12}, and RENO \cite{reno12}.
In global analysis, moreover, the values of $\sin^2\theta_{13}$ is 0.026(0.027) depending on the normal(inverted) neutrino mass ordering
\cite{global-ana}.
%%%

As the value of $\theta_B$ is correctly tuned, experimental value of
$\theta_{23}$ is realized. As there is no parameter to tune $\theta_{13}$, 
if $\theta_{13}\neq 0$ as is suggested by the recent experimental results, our model is excluded.
From the experimental constraints \cite{PDG},
\eqn{
\tan\theta_\nu=\frac{1}{\sqrt{2}},\quad
m^2_{\nu_2}-m^2_{\nu_1}=7.6\times 10^{-5}(\mbox{eV}^2),\quad
m^2_{\nu_2}-m^2_{\nu_3}=2.5\times 10^{-3}(\mbox{eV}^2), 
}
the phase $\phi$ is constrained by the condition
\eqn{
r\cos\phi =0.361,\quad r=\frac{\rho_2}{\rho_4}.
}
If we put the size of VEVs of Higgs fields as follow,
\eqn{
v_u=10,\quad v'_u=155.3,\quad v_d=2.0,\quad v'_d=77.8\quad (\mbox{GeV}),
\label{vev}
}
then from the charged lepton masses \cite{mass}
\eqn{
m^l_1=1.75\mbox{GeV},\quad  m^l_2=487\mbox{keV}, \quad m^l_3=103\mbox{MeV},
}
we can decide Yukawa coupling constants as follow,
\eqn{
Y^E_1=0.875,\quad Y^E_3=5.15\times 10^{-2},\quad Y^E_2=6.25\times 10^{-6}.
}
For the neutrinos, if we put
\eqn{
V=10^{12}\mbox{GeV},\quad Y^M_1=Y^M_3=1,
}
and assume $\phi=0$, all Yukawa coupling constants are decided as follow (see appendix)
\eqn{
\mbox{physical quantities}&:& \phi=\phi_2=0,\quad \phi_1=\pi , \no \\
&&m_{\nu_1}=5.240\times 10^{-2}\mbox{eV},\quad 
m_{\nu_2}=5.312\times 10^{-2}\mbox{eV},\quad m_{\nu_3}=1.795\times 10^{-2}\mbox{eV}
\no \\
\mbox{parameters}
&:&\delta=\frac{\pi}{2} ,\quad
\rho^2_2=1.795\times 10^{-2}\mbox{eV},\quad
\rho^2_3=15.51\times 10^{-2}\mbox{eV},\quad \rho^2_4=13.79\times 10^{-2}\mbox{eV} , \no \\
&&m^\nu_2=4.24\mbox{GeV},\quad m^\nu_3=12.45 \mbox{GeV},\quad m^\nu_4=11.74\mbox{GeV}, \no  \\
&&Y^N_2=0.424,\quad Y^N_3=0.080,\quad Y^N_4=1.17 .
}
For the lepton sector, there is no flavor changing process as same as quark sector
discussed above. Considering the interactions
\eqn{
{\cal L}_l
&=&Y^E_1\tau^c\L[l_\mu\L(\frac{H^D_2-H^D_1}{\sqrt{2}}\R)
+l_\tau\L(\frac{H^D_1+H^D_2}{\sqrt{2}}\R)\R]
-Y^E_2H^D_3e^cl_e \no \\
&+&Y^E_3\mu^c\L[l_\mu\L(\frac{H^D_1+H^D_2}{\sqrt{2}}\R)
+l_\tau\L(\frac{H^D_1-H^D_2}{\sqrt{2}}\R)\R],
}
as e does not couple to $\mu$ and $\tau$,
$\mu\to e\gamma$ and $\tau\to e\gamma$ processes  are forbidden.
Further more, due to the unbroken $S_2$ symmetry such as
$l_\mu\to-l_\mu,\mu^c\to-\mu^c,(H^D_1,H^D_2)\to(H^D_2,H^D_1)$,
$\tau\to\mu\gamma$ process is also forbidden.
This conclusion is not modified  even in the case of $\theta_B\neq \frac{\pi}{4}$. 
If there is small deviation from $\frac{\pi}{4}$ in $\theta_B$, under the $S_2$ translation 
$H^D_1\to H^D_1\cos2\theta_B+H^D_2\sin2\theta_B, H^D_2\to H^D_1\sin2\theta_B-H^D_2\cos2\theta_B$,
$H^D_1c_B+H^D_2s_B$ behaves like $S_2$-even field and $H^D_1s_B-H^D_2c_B$ behaves like $S_2$-odd field.
Note that the $S_4$ flavor symmetry does not help to solve SUSY flavor problem in our model 
because the quarks are assigned in $S_4$-singlet.
If the gaugino mass parameters are much larger than soft scalar masses, this problem may be solved.

\subsection{g-quark sector}

As the masses of g-quarks are given by
\eqn{
W_G=kS_3(G_1G^c_1+G_2G^c_2+G_3G^c_3),
}
the g-quark mass matrix  is proportional to unit matrix.
For the scalar g-quarks, as the contribution from 
soft flavor breaking term should be added, the degeneracy of masses may be broken.
However, such effects can be assumed as perturbation.
It is thought that dark matter does not have strong interaction,
g-quark should not be stable if the reheating temperature is higher than g-quark mass.
Under the symmetry defined in Table 1 and Table 2,
g-quarks can decay through the non-renormalizable terms as follows,
\eqn{
W_B&=&\frac{1}{M^2_P}Y^{QQG}_{ab}\Phi_3[G_1\Phi^c_1+G_2\Phi^c_2+G_3\Phi^c_3]Q_aQ_b \no \\
&+&\frac{1}{M^2_P}Y^{UDG}_{ab}\Phi_3[G^c_1\Phi^c_1+G^c_2\Phi^c_2+G^c_3\Phi^c_3]U^c_aD^c_b \no \\
&+&\frac{1}{M^2_P}Y^{EUG}_a\Phi_3[G_1\Phi^c_1+G_2\Phi^c_2+G_3\Phi^c_3]E^c_2U^c_a \no \\
&+&\frac{1}{M^2_P}Y^{L_sQG}_a\Phi_3[G^c_1\Phi^c_1+G^c_2\Phi^c_2+G^c_3\Phi^c_3]L_3Q_a \no \\
&+&\frac{1}{M^2_P}Y^{L_dQG}_a\Phi_3[\sqrt{3}L_1(G^c_2\Phi^c_2-G^c_3\Phi^c_3)
+L_2(G^c_2\Phi^c_2+G^c_3\Phi^c_3-2G^c_1\Phi^c_1)]Q_a.
}
Here we estimate the g-quark life time by the $Y^{EUG}$ interaction,
under the assumption  that only the decay to right-handed slepton is kinematically allowed,  
because it is natural to expect that the right-handed sleptons 
are lighter than the squarks and the left-handed sleptons in the result of
running  based on renormalization group equation.
If we put
\eqn{
Y^{EUG}_1=Y^{EUG}_2=Y^{EUG}_3=1,
}
then the interaction is given by
\eqn{
{\cal L}_g=\frac{(A^{EUG}_{RF})_SV^2}{\sqrt{3}M^2_P}E^c_2g_1(u^c_1+u^c_2+u^c_3),
}
where $E^c_2$ is right handed selectron, and the renormalization factor $(A^{EUG}_{RF})_S$
is evaluated by the RGE
\eqn{
(4\pi)\frac{d\ln Y^{EUG}_a}{d\ln\mu}=-\frac{16}{3}\alpha_s ,
}
and given by
\eqn{
(A^{EUG}_{RF})_S&=&\L(\frac{M_P}{M_Z}\R)^{4\alpha_s/3\pi}
=\L(\frac{2.43\times 10^{18}}{91}\R)^{0.05008}=6.647,
}
where only QCD correction is accounted. This approximation is not bad
because the beta function of the coupling constant of strong interaction $g_s$ vanishes at 1-loop level
in our model, which makes the contribution of $\alpha_s$ dominant in  the RGE of $Y^{EUG}$.

Using the interaction Eq.(49), we calculate the g-quark decay width.
For simplicity, we assume $u^c_a$ in Eq.(49) are mass eigenstates.
Requiring the life time of g-quark is shorter than 0.1 sec
(otherwise the success of BBN is spoiled \cite{g-lifetime}) as follow  
\eqn{
\Gamma(g_1)=3\L(\frac{(A^{EUG}_{RF})_SV^2}{3M^2_P}\R)^2\frac{M_g}{16\pi}
>\frac{1}{0.1\ \mbox{sec}} , %=6.582\times 10^{-24}GeV
}
we get
\eqn{
\frac{M_g}{1000 \mbox{GeV}}\L(\frac{V}{M_P}\R)^4>2.25 \times 10^{-26},
}
which bounds the VEV size of flavon from below.

Finally, using the interaction
$Y^{QQG}-Y^{EUG}$, we estimate the proton decay width.
Integrating out the scalar g-quarks, 
we get the effective four-Fermi interactions as follows
\eqn{
{\cal L}_{p\to e^+\pi^0}=\frac{V^4}{M^4_PM^2_G}
Y^{QQG}_{ab}Y^{EUG}_cA_{RF}\bar{q}_a\bar{q}_bu^c_ce^c,
}
where
\eqn{
A_{RF}&=&(A^{EUG}_{RF})_S(A^{QQG}_{RF})_S(A_{RF})_L \quad \cite{p-RGE},
}
and the renormalization factor $(A^{QQG}_{RF})_S$ is evaluated by the REG
\eqn{
(4\pi)\frac{d\ln Y^{QQG}_a}{d\ln\mu}=-\frac{24}{3}\alpha_s,
}
and given by
\eqn{
(A^{QQG}_{RF})_S&=&\L(\frac{M_P}{M_Z}\R)^{2\alpha_s/\pi}
=\L(\frac{2.43\times 10^{18}}{91}\R)^{0.07512}=17.139.
}
As the long distant part of renormalization factor is given by 
\eqn{
(A_{RF})_L=\L(\frac{\alpha_s(1\mbox{GeV})}{\alpha_s(m_b)}\R)^{6/25}
\L(\frac{\alpha_s(m_b)}{\alpha_s(M_Z)}\R)^{6/23}=1.4  \quad \cite{p-decay},
}
we get
\eqn{
A_{RF}=159.5.
}
In the quark mass basis, Eq.(54) is rewritten as follow
\eqn{
{\cal L}_{p\to e^+\pi^0}&=&\frac{V^4}{M^4_PM^2_G}
\L[2(L^T_uY^{UDG}L_u)_{ab}(Y^{EUG}R_u)_c\R]A_{RF}\bar{u}'_a\bar{d}'_b(u^c_c)'e^c, \\
&&\bar{u}=L_u\bar{u}',\quad \bar{d}=L_d\bar{d}',\quad u^c=R_u(u^c)'.
}
For simplicity, we put
\eqn{
\L[2(L^T_uY^{UDG}L_u)_{11}(Y^{EUG}R_u)_1\R]=1,
}
then the proton decay width is given by
\eqn{
\Gamma(p\to \pi^0+e^+)&=&\frac{m_p}{64\pi f^2_\pi}
\L[\L(\frac{V}{M_P}\R)^4\frac{A_{RF}}{M^2_G}\R]^2(1+F+D)^2
\L(1-\frac{m^2_{\pi^0}}{m^2_p}\R)^2\alpha^2_p \quad \cite{PDform}.
}
If we put
\eqn{
&&F=0.47,\quad D=0.80,\quad \alpha_p=-0.012\ \mbox{GeV}^3,\quad f_\pi=130\ \mbox{MeV}, \quad  
\cite{chiral}\no \\
&&m_{\pi^0}=135\ \mbox{MeV},\quad m_p=940\ \mbox{MeV}, \quad \cite{PDG}
}
then we get
\eqn{
\Gamma(p\to \pi^0+e^+)&=&(5.01 \times 10^{-12}\ \mbox{GeV})
\L[\L(\frac{V}{M_P}\R)^4\frac{(1000\ \mbox{GeV})^2}{M^2_G}\R]^2 .
}
From the experimental bound $\tau(p\to \pi^0+e^+)>1600\times 10^{30}[\mbox{years}]$ \cite{PDG},
the VEV size of flavon is bounded from above  as follow
\eqn{
\L[\L(\frac{V}{M_P}\R)^4\L(\frac{1000\ \mbox{GeV}}{M_G}\R)^2\R]^2
< 2.60\times 10^{-54}.
}
Hereafter we assume the approximation $M_g=M_G$ is held for simplicity.
From Eq.(53) and Eq.(66), the allowed region for $V$ is given by (see Fig.1)
\eqn{
2.25\times 10^{-26}\L(\frac{1000 \mbox{GeV}}{M_G}\R)<\L(\frac{V}{M_P}\R)^4
<1.61\times 10^{-27}\L(\frac{M_G}{1000\ \mbox{GeV}}\R)^2 .
}
This inequality holds when the mass bound,
\eqn{
M_G> 2.41 \mbox{TeV},
}
is satisfied.

%%%%%%%%%%%%%%%%%%%%%%%%%%%%%%%%%%%%%%
\begin{figure}[ht]
\unitlength=1mm \hspace{3cm}
\begin{picture}(70,70)
\includegraphics[height=6cm,width=10cm]{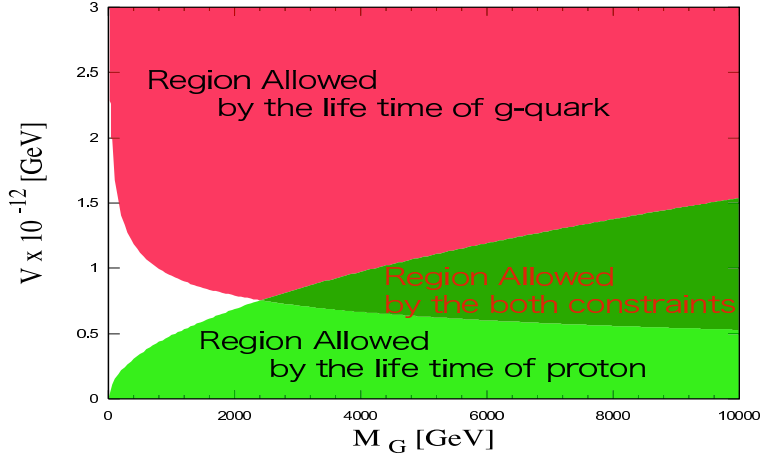}
\end{picture}
\caption{ $M_G$ versus $V$: The pink region comes from the constraint of the life time of the g-quark, which should be less than 0.1 sec.  The green region comes from the constraint of the proton stability. The black region is allowed by the both constraints. The heavier of $M_G$, the wider the allowed region is.}
\label{proton}
\end{figure}
%%%%%%%%%%%%%%%%%%%%%%%%%%%%%%%%%%%%%%%

Before ending this section, we discuss the unsatisfactory point of this model.
Considering the mass spectrum of the quarks and charged leptons,
it is expected that the trilinear  coupling of first generation is multi suppressed 
by the suppression mechanism of Yukawa couplings and $S_4$ symmetry.
If it is true, as the proton decay width accommodates another suppression factor,
the condition Eq.(62) is never satisfied and
the experimental verification of proton decay seems to be impossible.
Therefore the bounds Eq.(66)-(68) should not be taken seriously.
In the realistic model, the suppression by the gauge non-singlet flavon may be too strong.
To improve this point, we modify the flavon sector in next section.

\section{$S_4\times Z_4\times Z_9$ flavor symmetric model}

In this section we introduce Froggatt-Nielsen mechanism to explain Yukawa hierarchy \cite{fn},
and the flavon sector is modified as follows.
To realize $O(10^2)$ hierarchy, we introduce $Z_9$ symmetry and add gauge and $S_4$ singlet $X$
as Froggatt-Nielsen (FN) flavon. 
To weaken the over suppression of trilinear  terms, we replace the flavon $\Phi^c_a$ 
by $S_4$ singlet $\Phi^c$ and gauge singlet $T$ which is assigned to $S_4$ triplet.
To forbid renormalizable terms of $T$, the $Z_2$ symmetry is replaced by $Z_4$.
The flavor representations of superfields are given in Table 4. 

\begin{table}[htbp]
\begin{center}
\begin{tabular}{|c|c|c|c|c|c|c|c|c|c|c|c|c|c|c|}
\hline
        &$Q_1$    &$Q_2$    &$Q_3$     &$U^c_1$  &$U^c_2$  &$U^c_3$   &$D^c_1$  &$D^c_2$   &$D^c_3$ 
        &$E^c_1$  &$E^c_2$  &$E^c_3$   &$L_i$    &$L_3$ \\
      \hline
$S_4$   &${\bf 1}$&${\bf 1}$& ${\bf 1}$&${\bf 1}$&${\bf 1}$&${\bf 1}$ &${\bf 1}$&${\bf 1}$ &${\bf 1}$
        &${\bf 1}$&${\bf 1}$&${\bf 1'}$&${\bf 2}$&${\bf 1}$ \\
      \hline
$Z_4$   &$1/2$    &$1/2$    &$1/2$     &$1/2$    &$1/2$    &$1/2$     &$1/2$    &$1/2$     &$1/2$ 
        &$0$      &$1/2$    &$0$       &$1/2$    &$1/2$ \\
      \hline
$Z_9$   &$1/9$    &$1/9$    &$0$       &$2/9$    &$1/9$    &$0$       &$2/9$    &$1/9$     &$1/9$ 
        &$0$      &$2/9$    &$1/9$     &$0$      &$1/9$ \\    
      \hline
      \hline
        &$N^c_i$  &$N^c_3$  &$H^U_i$   &$H^U_3$  &$H^D_i$  &$H^D_3$   &$S_i$    &$S_3$       
        &$G_a$    &$G^c_a$  &$T_a$     &$\Phi$   &$\Phi^c$ &$X$  \\
      \hline
$S_4$   &${\bf 2}$&${\bf 1}$&${\bf 2}$ &${\bf 1}$&${\bf 2}$ &${\bf 1}$&${\bf 2}$&${\bf 1}$
        &${\bf 3}$&${\bf 3}$ &${\bf 3}$&${\bf 1}$&${\bf 1'}$&${\bf 1}$\\
      \hline      
$Z_4$   &$0$      &$1/2$    &$1/2$     &$0  $    &$1/2$     &$0$      &$1/2$    &$0$         
        &$1/4$    &$3/4$     &$1/4$    &$0$      &$1/2$     &$0$\\
      \hline
$Z_9$   &$0$      &$0$      &$0$       &$0$      &$0$       &$0$      &$0$      &$0$          
        &$0$      &$0$       &$0$      &$0$      &$0$       &$8/9$\\ 
      \hline
\end{tabular}
\end{center}
\caption{$S_4\times Z_4\times Z_9$ assignment of superfields
(Where the index $i$ of the $S_4$ doublets runs $i=1,2$,
and the index $a$ of the $S_4$ triplets runs $a=1,2,3$.)}
\end{table}

\subsection{Flavon sector}

The leading terms of flavons are given as follows,
\eqn{
W_F&=&W_T+W_\Phi+W_X,\\
W_T&=&\frac{1}{M_P}\L[\frac14 Y^T_1(T^4_1+T^4_2+T^4_3)
+\frac12 Y^T_2(T^2_1T^2_2+T^2_1T^2_3+T^2_2T^2_3)\R] ,\\
W_X&=&\frac{1}{6M^6_P}X^9, \\
W_\Phi&=&\frac{1}{2M_P}Y^\Phi(\Phi\Phi^c)^2.
}
The VEV size of gauge non-singlet is estimated by Eq.(24).
Now we change the value of $V=\L<\Phi\R>$ to $10^{11}$ GeV which is
given by naive estimation  when we put $Y^\Phi\sim 1$.
This affects neutrino Yukawa couplings given in Eq.(44) as follows
\eqn{
Y^N_2\to \frac{Y^N_2}{\sqrt{10}}=0.134,\quad
Y^N_3\to \frac{Y^N_3}{\sqrt{10}}=0.025,\quad
Y^N_4\to \frac{Y^N_4}{\sqrt{10}}=0.370.
}
The VEV size of FN flavon is estimated as follow
\eqn{
\epsilon=\L(\frac{\L<X\R>}{M_P}\R)&\sim&\L(\frac{m_{SUSY}}{M_P}\R)^{\frac{1}{7}}
\sim 10^{-2}.
}
$S_4$ symmetric part of potential of $T$ is given by
\eqn{
V_T&=&-m^2(|T_1|^2+|T_2|^2+|T_3|^2) \no \\
&-&\frac{1}{M_P}\L[\frac14 B^T_1(T^4_1+T^4_2+T^4_3)
+\frac12 B^T_2(T^2_1T^2_2+T^2_1T^2_3+T^2_2T^2_3)\R] \no \\
&+&\frac{1}{M^2_P}\L[|Y^T_1T^3_1+Y^T_2T_1(T^2_2+T^2_3)|^2
+|Y^T_1T^3_2+Y^T_2T_2(T^2_1+T^2_3)|^2 \R. \no \\
&+&\L. |Y^T_1T^3_3+Y^T_2T_3(T^2_1+T^2_2)|^2\R],
}
which has minimum in the VEV direction given by
\eqn{
T_a=\frac{V_T}{\sqrt{3}}(1,1,1).
}
As same as the gauge non-singlet model, we can assume
flavor breaking term as perturbation.
As the size of $V_T$ is at the same order as $V$, we put
\eqn{
\frac{V_T}{M_P}=10^{-8}.
}

\subsection{Quark and Lepton sector}

Due to the $Z_9$ symmetry, the effective Yukawa coupling constants accommodate  power of $\epsilon$
through the superpotential 
\eqn{
W_Q=\sum_{ij}\L(\frac{X}{M_P}\R)^{9(q_i+u_j)}\L(Y^U_{ij}\R)_0H^U_3Q_iU^c_j
+\sum_{ij}\L(\frac{X}{M_P}\R)^{9(q_i+d_j)}\L(Y^D_{ij}\R)_0H^D_3Q_iD^c_j,
}
where $q_i,u_i,d_i$ are $Z_9$ charge of $Q_i,U^c_i,D^c_i$ respectively.
As the results, the mass matrices of quarks are given as follows,
\eqn{
M_u&\sim&\Mat3{\epsilon^{3}}{\epsilon^{2}}{\epsilon}
{\epsilon^{3}}{\epsilon^{2}}{\epsilon}
{\epsilon^{2}}{\epsilon}{1}
\sim \Mat3{1}{1}{\epsilon}{1}{1}{\epsilon}{\epsilon}{\epsilon}{1}
\mbox{diag}(\epsilon^3,\epsilon^2,1) 
\Mat3{1}{\epsilon}{\epsilon^2}{\epsilon}{1}{\epsilon}{\epsilon^2}{\epsilon}{1}v'_u
=L_u\mbox{diag}(m_{u,c,t})R^\dagger_u, \\
M_d&\sim&\Mat3{\epsilon^{3}}{\epsilon^{2}}{\epsilon^{2}}
{\epsilon^{3}}{\epsilon^{2}}{\epsilon^{2}}
{\epsilon^{2}}{\epsilon}{\epsilon}
\sim \Mat3{1}{1}{\epsilon}{1}{1}{\epsilon}{\epsilon}{\epsilon}{1}
\mbox{diag}(\epsilon^3,\epsilon^2,\epsilon) 
\Mat3{1}{\epsilon}{\epsilon}{\epsilon}{1}{1}{\epsilon}{1}{1}v'_d
=L_d\mbox{diag}(m_{d,s,b})R^\dagger_d,
}
from which we get Cabbibo-Kobayashi-Maskawa matrix
\eqn{
V_{CKM}=L^\dagger_u L_d \sim \Mat3{1}{1}{\epsilon}{1}{1}{\epsilon}{\epsilon}{\epsilon}{1} , 
}
and quark masses divided by experimental values respectively,
\eqn{
&&\frac{m_u}{(m_u)_{\mbox{exp}}}=\frac{(A^Y_{RF})_S\epsilon^3 v'_u}{1.3\times 10^{-3}}=0.79 ,\quad
\frac{m_c}{(m_c)_{\mbox{exp}}}=\frac{(A^Y_{RF})_S\epsilon^2 v'_u}{0.624}=0.17 ,\quad
\frac{m_t}{(m_t)_{\mbox{exp}}}=\frac{v'_u}{173}=0.90 , \no \\
&&\frac{m_d}{(m_d)_{\mbox{exp}}}=\frac{(A^Y_{RF})_S\epsilon^3 v'_d}{2.9\times 10^{-3}}=0.18 ,\quad
\frac{m_s}{(m_s)_{\mbox{exp}}}=\frac{(A^Y_{RF})_S\epsilon^2 v'_d}{0.055}=0.94 ,\quad
\frac{m_b}{(m_b)_{\mbox{exp}}}=\frac{(A^Y_{RF})_S\epsilon v'_d}{2.89}=1.8 , \no \\
}
where $\epsilon=0.01, v'_u=155.3\mbox{GeV},v'_d=77.8\mbox{GeV}, (A^Y_{RF})_S=(A^{EUG}_{RF})_S=6.647$
 are used. The renormalization factor of top-Yukawa coupling  is neglected because it has
infrared  quasi-fixed point.
For the lepton sector, the Yukawa coupling constants divided by
required values given in Eq.(42) and Eq.(73) are given by
\eqn{
&&\frac{Y^E_1}{(Y^E_1)_{\mbox{exp}}}=\frac{1}{0.875}=1.1,\quad
\frac{Y^E_3}{(Y^E_3)_{\mbox{exp}}}=\frac{\epsilon}{5.15\times 10^{-2}}=0.19,\quad
\frac{Y^E_2}{(Y^E_2)_{\mbox{exp}}}=\frac{\epsilon^3}{6.25\times 10^{-6}}=0.16, \no \\
&&\frac{Y^N_2}{(Y^N_2)_{\mbox{exp}}}=\frac{1}{0.134}=7.5,\quad
\frac{Y^N_3}{(Y^N_3)_{\mbox{exp}}}=\frac{\epsilon}{0.025}=0.40,\quad
\frac{Y^N_4}{(Y^N_4)_{\mbox{exp}}}=\frac{1}{0.370}=2.7.
}
Where we used running masses of quarks and charged leptons \cite{mass}:
\eqn{
\begin{tabular}{lll}
$m_u(m_Z)=1.28^{+0.50}_{-0.39} (\mbox{MeV})$, &
$m_c(m_Z)=624\pm 83 (\mbox{MeV})$, &
$m_t(m_Z)=172.5\pm 3.0 (\mbox{GeV})$, \\
$m_d(m_Z)=2.91^{+1.24}_{-1.20} (\mbox{MeV})$, &
$m_s(m_Z)=55^{+16}_{-15} (\mbox{MeV})$, &
$m_b(m_Z)=2.89\pm 0.09 (\mbox{GeV})$, \\
$m_e(m_Z)=0.48657 (\mbox{MeV})$, &
$m_\mu(m_Z)=102.72 (\mbox{MeV})$, &
$m_\tau(m_Z)=1746 (\mbox{MeV})$.
\end{tabular}
}
The discrepancies  between the estimated values and experimental  values 
in Eq.(82) and Eq.(83) are easily  recovered by multiplying $O(1)$ coefficients  $(Y)_0$.
Here we assume $O(1)$ means $0.5<O(1)<5$, therefore $Y^N_2$ is out of this range.
However our fitting is not totally wrong.
CKM matrix
\eqn{
(V_{CKM})_{\mbox{exp}}\simeq
\Mat3{1}{0.23}{0.4\times 10^{-2}}
{0.23}{1}{4.1\times 10^{-2}}
{0.8\times10^{-2}}{3.9\times10^{-2}}{1}, \quad \cite{PDG}
}
is also recovered  from Eq.(81). Therefore our procedure works well in quark and lepton sectors.

\subsection{g-quark sector}

The leading terms of single g-quark interactions are given by
\eqn{
W_B&=&\frac{1}{M_P}\L(\frac{X}{M_P}\R)^{9(u_a+d_a)}(Y^{UDG}_{ab})_0
[T_1G^c_1+T_2G^c_2+T_3G^c_3]U^c_aD^c_b \no \\
&+&\frac{1}{M_P}\L(\frac{X}{M_P}\R)^{9q_a+1}(Y^{QL_sG}_a)_0
[T_1G^c_1+T_2G^c_2+T_3G^c_3]Q_aL_3 \no \\
&+&\frac{1}{M_P}\L(\frac{X}{M_P}\R)^{9q_a}(Y^{QL_dG}_a)_0
[\sqrt{3}(T_2G^c_2-T_3G^c_3)L_1+(T_2G^c_2+T_3G^c_3-2T_1G^c_1)L_2]Q_a,
}
where the contribution from $E^c_1\supset \tau^c$ is omitted because $p\to\tau^+X$
is impossible. Note that $Y^{QQG},Y^{EUG}$ are suppressed by $(V_T/M_P)^3$. 
In the quark mass basis, trilinear  coupling matrix and vectors  are given as follows, 
\eqn{
(R^T_uY^{UDG}R_d)_{ab}&\sim&
\Mat3{1}{\epsilon}{\epsilon^2}{\epsilon}{1}{\epsilon}{\epsilon^2}{\epsilon}{1}
\Mat3{\epsilon^4}{\epsilon^3}{\epsilon^3}
{\epsilon^3}{\epsilon^2}{\epsilon^2}
{\epsilon^2}{\epsilon}{\epsilon}
\Mat3{1}{\epsilon}{\epsilon}{\epsilon}{1}{1}{\epsilon}{1}{1}
\sim \Mat3{\epsilon^4}{\epsilon^3}{\epsilon^3}
{\epsilon^3}{\epsilon^2}{\epsilon^2}
{\epsilon^2}{\epsilon}{\epsilon} , \\
(Y^{QL_sG}L_u)_a&\sim&\epsilon (\epsilon,\epsilon,1)
\Mat3{1}{1}{\epsilon}{1}{1}{\epsilon}{\epsilon}{\epsilon}{1}
\sim(\epsilon^2,\epsilon^2,\epsilon), \\
(Y^{QL_dG}L_u)_a&\sim&(\epsilon,\epsilon,1)
\Mat3{1}{1}{\epsilon}{1}{1}{\epsilon}{\epsilon}{\epsilon}{1}
\sim(\epsilon,\epsilon,1).
}
As the coupling constants are large enough,  the problem of
long life time of g-quark is solved.
Integrating out the scalar g-quarks, 
would-be the largest contribution to proton decay is given by
\eqn{
{\cal L}_{p\to \mu^+K^0}&=&\frac{\epsilon^4}{M^2_PM^2_G}
(\bar{u}^c)'(\bar{s}^c)'u'[\sqrt{3}e_1(\L<T_2\R>^2-\L<T_3\R>^2)
+e_2(\L<T_2\R>^2+\L<T_3\R>^2-2\L<T_1\R>^2)], 
}
where $e_1,e_2$ are linear combinations of $\mu$ and $\tau$.
Interestingly, this interaction vanishes in the VEV direction given in Eq.(76).
This means the contributions from three scalar g-quarks are canceled.
Therefore the dominant contribution to proton decay is given by 
\eqn{
{\cal L}_{p\to e^+K^0}
=\frac{\epsilon^5V^2_T}{M^2_PM^2_G}A_{RF}(\bar{u}^c)'(\bar{s}^c)'u'e,
}
from which we get
\eqn{
\Gamma(p\to e^+ +K^0)&=&\frac{m_p}{32\pi f^2_\pi}
\L[\epsilon^5\L(\frac{V_T}{M_P}\R)^2\frac{A_{RF}}{M^2_G}\R]^2
\L[-1+\frac{m_N}{m_{B'}}(F-D)\R]^2\L(1-\frac{m^2_{K^0}}{m^2_p}\R)^2\alpha^2_p.
}
Substituting 
\eqn{
\epsilon=0.01,\quad
m_N=m_p=940\mbox{MeV},\quad m_{B'}=\frac{m_\Lambda+m_\Sigma}{2}=1150\mbox{MeV},\quad
m_{K^0}=498\mbox{MeV},\quad \cite{PDG}
}
and Eq.(59), Eq.(64) and Eq.(77), we get
\eqn{
\Gamma(p\to e^+ +K^0)&=&1.69\times 10^{-64}\L(\frac{1000\mbox{GeV}}{M_G}\R)^4\mbox{GeV}.
}
For the experimental bound $\tau(p\to e^+ +K^0)>150 \times 10^{30}[\mbox{years}]$ \cite{PDG},
mass bound as follows,
\eqn{
M_G&>&1.0\mbox{TeV}
}
must be satisfied. The experimental bound 
$\tau(p\to \bar{\nu} +K^+)>670 \times 10^{30}[\mbox{years}]$ \cite{PDG} for the operator
\eqn{
{\cal L}_{p\to \bar{\nu}K^+}
=\frac{\epsilon^5V^2_T}{M^2_PM^2_G}A_{RF}(\bar{u}^c)'(\bar{s}^c)'d'\nu,
}
gives weaker mass bound as follows,
\eqn{
\Gamma(p\to \bar{\nu} +K^+)&=&\frac{m_p}{32\pi f^2_\pi}
\L[\epsilon^5\L(\frac{V_T}{M_P}\R)^2\frac{A_{RF}}{M^2_G}\R]^2
\L[\frac23 \frac{m_N}{m_{B'}}D\R]^2\L(1-\frac{m^2_{K^0}}{m^2_p}\R)^2\alpha^2_p \no \\
&=&0.20\times 10^{-64}\L(\frac{1000\mbox{GeV}}{M_G}\R)^4\mbox{GeV}  , \\
M_G&>&0.9\mbox{TeV}.
}
Note that these bounds should not be taken seriously, 
because there is $O(10)$ ambiguity coming from SUSY breaking parameter
in $V_T$, this constraint also has such ambiguity. 
The important point is that
we can expect the experimental observation of proton decay for TeV scale g-quark in near future
\footnote{The recent experimental mass bound for $Z'$ gauge boson 
$m_{Z'}\simeq 0.515v'_s>1.52$ TeV \cite{newZp}
gives g-quark mass bound $m_g=kv'_s>3$ TeV for $g_X=g_Y,k=1$.}.

Finally we give a short comment about flavor breaking effects on cancellation.
Due to the perturbation from flavor breaking squared mass terms,
scalar g-quark squared mass and squared VEVs of flavons receive
$O(m^2_B/m^2_{SUSY})$ contaminations, where $m^2_B$ is flavor breaking squared mass.
As the results, the cancellation is spoiled.
However, if flavor breaking terms are small enough to satisfy the condition
\eqn{
r_B=\frac{m^2_B}{m^2_{SUSY}}< \epsilon,
}
the cancellation mechanism works effectively. 
In the  opposite  case of $r_B>\epsilon$, $p\to \mu^+K^0$ dominates the proton decay width.
Therefore the size of $r_B$ affects  proton decay channels significantly.
There is interesting  correlation through $r_B$ between proton decay channel and
degree of degeneracy of scalar g-quark masses.
Note that too small $r_B$ causes the appearance of pseudo-Nambu-Goldstone boson (pNGB). 
However this may be not a serious problem, because  if there is 
the VEV hierarchy such as
$v_s\gg v_u,v_d$,  which makes $S_i$ dominant in pNGB, then
the interactions of this pNGB with quarks, leptons and weak bosons
are very weak.

\section{Dirac neutrino model}

As is shown in previous section,  the stability of proton is realized by 
strong suppression factor of $\L<T\R>/M_P\sim 10^{-8}$.
It is not unnatural to expect that the small neutrino mass is realized by same mechanism.
In this section, we construct Dirac neutrino model based on $S_4$ flavor symmetry.
We eliminate $\Phi,\Phi^c$ and change the flavor assignment as given in Table 5.
In this model, as we can not break $U(1)_Z$ gauge symmetry, only one extra U(1) gauge symmetry
is allowed to exist. As the RHNs do not have $U(1)_X$ charge, Majorana mass terms are not
forbidden by $U(1)_X$. Therefore we change the extra gauge symmetry to
one linear combination of two extra U(1) gauge
symmetries as defined by
\eqn{
X_\theta=X\cos\theta+Z\sin\theta,
} 
and assume RHN and $S$ have non-zero charge of $X_\theta$. The value of $\theta$
does not affect our analysis.

\begin{table}[htbp]
\begin{center}
\begin{tabular}{|c|c|c|c|c|c|c|c|c|c|}
\hline
        &$Q_1$    &$Q_2$    &$Q_3$     &$U^c_1$  &$U^c_2$  &$U^c_3$   &$D^c_1$  &$D^c_2$   &$D^c_3$ \\
      \hline
$S_4$   &${\bf 1}$&${\bf 1}$& ${\bf 1}$&${\bf 1}$&${\bf 1}$&${\bf 1}$ &${\bf 1}$&${\bf 1}$ &${\bf 1}$\\
      \hline
$Z_4$   &$1/2$    &$1/2$    &$1/2$     &$1/2$    &$1/2$    &$1/2$     &$1/2$    &$1/2$     &$1/2$ \\
      \hline
$Z_9$   &$1/9$    &$1/9$    &$0$       &$2/9$    &$1/9$    &$0$       &$2/9$    &$1/9$     &$1/9$ \\    
      \hline
      \hline
        &$E^c_1$  &$E^c_2$  &$E^c_3$   &$L_i$    &$L_3$    &$H^U_i$   &$H^U_3$  &$H^D_i$   &$H^D_3$  \\
      \hline
$S_4$   &${\bf 1}$&${\bf 1}$&${\bf 1'}$&${\bf 2}$&${\bf 1}$&${\bf 2}$ &${\bf 1}$&${\bf 2}$ &${\bf 1}$ \\
      \hline      
$Z_4$   &$0$      &$1/2$    &$0$       &$1/2$    &$1/2$    &$1/2$     &$0  $    &$1/2$     &$0$ \\
      \hline
$Z_9$   &$0$      &$3/9$    &$1/9$     &$0$      &$0$      &$0$       &$0$      &$0$       &$0$ \\ 
      \hline
\end{tabular}
\begin{tabular}{|c|c|c|c|c|c|c|c|}      
      \hline   
        &$S_i$    &$S_3$    &$N^c_a$   &$G_a$    &$G^c_a$   &$T_a$    &$X$ \\
      \hline
$S_4$   &${\bf 2}$&${\bf 1}$&${\bf 3}$ &${\bf 3}$&${\bf 3}$ &${\bf 3}$&${\bf 1}$\\
      \hline
$Z_4$   &$1/2$    &$0$      &$3/4$     &$1/4$    &$3/4$     &$1/4$    &$0$ \\
      \hline
$Z_9$   &$0$      &$0$      &$2/9$     &$0$      &$0$       &$0$      &$8/9$ \\  
      \hline
\end{tabular}
\end{center}
\caption{$S_4\times Z_4\times Z_9$ assignment of superfields
(Where the index $i$ of the $S_4$ doublets runs $i=1,2$,
and the index $a$ of the $S_4$ triplets runs $a=1,2,3$.)}
\end{table}

As the quark and charged lepton Yukawa interactions are not modified, we
consider only Yukawa interaction of neutrino which is given by
\eqn{
W_N&=&\frac{X^2}{M^3_P}(Y^N_1)_0(T_1N^c_1+T_2N^c_2+T_3N^c_3)(H^U_1L_1+H^U_2L_2) \no \\
&-&\frac{X^2}{M^3_P}(Y^N_2)_0[\sqrt{3}(T_2N^c_2-T_3N^c_3)(H^U_1L_2+H^U_2L_1)
+(T_2N^c_2+T_3N^c_3-2T_1N^c_1)(H^U_1L_1-H^U_2L_2)] \no \\
&-&\frac{X^2}{M^3_P}(Y^N_3)_0[\sqrt{3}(T_2N^c_2-T_3N^c_3)H^U_1
+(T_2N^c_2+T_3N^c_3-2T_1N^c_1)H^U_2]L_3.
}
Substituting the VEVs given in Eq.(74), Eq.(76) and Eq.(77) for $X$ and $T_a$,
we get the effective superpotential as follow
\eqn{
W_N&=&Y^N_1(H^U_1L_1+H^U_2L_2)(N^c_1+N^c_2+N^c_3) \no \\
&+&Y^N_2[\sqrt{3}(N^c_3-N^c_2)(H^U_1L_2+H^U_2L_1)
+(2N^c_1-N^c_2-N^c_3)(H^U_1L_1-H^U_2L_2)] \no \\
&+&Y^N_3L_3[\sqrt{3}(N^c_3-N^c_2)H^U_1+(2N^c_1-N^c_2-N^c_3)H^U_2] , \\
&&Y^N_{1,2,3}=(Y^N_{1,2,3})_0\epsilon^2\L(\frac{V_T}{\sqrt{3}M_P}\R)\sim O(10^{-12}), 
}
from which Dirac neutrino mass matrix is given by
\eqn{
M_D&=&\Mat3{(m_1+2m_2)c_u}{m_1c_u-m_2(c_u+\sqrt{3}s_u)}{m_1c_u-m_2(c_u-\sqrt{3}s_u)}
{(m_1-2m_2)s_u}{m_1s_u+m_2(s_u-\sqrt{3}c_u)}{m_1s_u+m_2(s_u+\sqrt{3}c_u)}
{2m_3s_u}{-m_3(s_u+\sqrt{3}c_u)}{m_3(-s_u+\sqrt{3}c_u)} , \\
&&m_1=Y^N_1v_u,\quad m_2=Y^N_2v_u,\quad m_3=Y^N_3v_u,
}
where we can define $m_1$ and $m_3$ as real and non-negative and $m_2$ as complex
without loss of generality. 
In the VEV direction $\theta_u=\theta_d=\theta_B$, two large angles of
charged lepton and neutrino mixing matrix are canceled  and  make it difficult
to realize two large mixing angles of MNS matrix.
Therefore we select the condition Eq.(17) and put $\theta_d=0$ and 
$\theta_u=\frac{\pi}{4}$ 
by hand (Note that $V_{MNS}$ depends on $\theta_u,\theta_d$ only through $\theta_u-\theta_d$.
To realize maximal mixing of $\theta_{23}$, we tune $\theta_u-\theta_d=\frac{\pi}{4}$.), 
then the charged lepton mass matrix is given by
\eqn{
M_l&=&
\Mat3{m^e_1}{0}{0}
{0}{0}{m^e_3}
{0}{m^e_2}{0}, 
}
from which we get
\eqn{
V^T_lM_lM^T_lV_l&=&\mbox{diag}((m^e_2)^2,(m^e_3)^2,(m^e_1)^2)
=\mbox{diag}(m^2_e,m^2_\mu,m^2_\tau),\quad
V_l=\Mat3{0}{0}{1}{0}{1}{0}{1}{0}{0}.
}
To realize experimental  results, the conditions given as follows
\eqn{
V^\dagger_\nu M^*_DM^T_DV_\nu &=&diag(m^2_{\nu_1},m^2_{\nu_2},m^2_{\nu_3})=M^2_{\mbox{diag}}, \\
V_\nu&=&V_lV_{MNS}=\frac{1}{\sqrt{6}}V_l\Mat3{\sqrt{2}}{0}{0}{0}{1}{1}{0}{-1}{1}
\Mat3{1}{0}{-\lambda^*}{0}{1}{0}{\lambda}{0}{1} 
\Mat3{\sqrt{2}}{1}{0}{-1}{\sqrt{2}}{0}{0}{0}{\sqrt{3}}, \\
M^*_DM^T_D&=&V_\nu M^2_{\mbox{diag}} V^\dagger_\nu
=\mbox{diag}(1,1,1)m^2_{\nu_2}+\Delta m^2_{32}V_\nu \mbox{diag}(-r_\nu,0,1)V^\dagger_\nu, \\
r_\nu&=&\frac{\Delta m^2_{21}}{\Delta m^2_{32}},
}
must be satisfied. Eq.(110) is rewritten as follow
\eqn{
&&\Mat3{(3/2)m^2_1+6|m_2|^2-m^2_{\nu_2}}{(3/2)m^2_1}{6m^*_2m_3}
{(3/2)m^2_1}{(3/2)m^2_1+6|m_2|^2-m^2_{\nu_2}}{0}
{6m_2m_3}{0}{6m^2_3-m^2_{\nu_2}} \no \\
&=&\frac{\Delta m^2_{32}}{6}
\Mat3{3-r_\nu}{3+r_\nu}{-2r_\nu-3\sqrt{2}\lambda}
{3+r_\nu}{3-r_\nu}{2r_\nu-3\sqrt{2}\lambda}
{-2r_\nu-3\sqrt{2}\lambda^*}{2r_\nu-3\sqrt{2}\lambda^*}{-4r_\nu},
}
where $O(\lambda^2)$ terms are neglected. From this equation, we get 
\eqn{
&&\sin\theta_{13}=\lambda=\lambda^*=\frac{\sqrt{2}}{3}r_\nu
=\frac{\sqrt{2}}{3}\frac{\Delta m^2_{21}}{\Delta m^2_{32}}
=0.014\quad (\theta_{13}=0.8^\circ) , \\
&&m_1=0.029\mbox{eV} ,\quad
m_2=-0.0033\mbox{eV},\quad
m_3=0.0025\mbox{eV}  \no \\
&&m_{\nu_1}=0.0035\mbox{eV},\quad m_{\nu_2}=0.0094\mbox{eV},\quad m_{\nu_3}=0.051\mbox{eV} , \\
&&Y^N_1=2.9\times 10^{-12},\quad Y^N_2=-0.33\times 10^{-12},\quad 
Y^N_3=0.25\times 10^{-12}.
}
The small discrepancies between Eq.(103) and Eq.(115) are recovered 
by multiplying O(1) coefficients $(Y^N)_0$.

By the modification of $Z_9$ charge of $L_3$, 
the proton decay width is dominated by $p\to e^+K^0$ given in Eq.(91) 
because suppression factor is reduced from $\epsilon^5$ to $\epsilon^4$.
From the experimental  bound, we get
\eqn{
M_G>\frac{1.0\mbox{TeV}}{\sqrt{\epsilon}}=10\mbox{TeV}.
}
Note that $Y^{NDG}$ is suppressed by $(V_T/M_P)^2$.

\section{Conclusion}

We have considered  proton stability based on $S_4$ symmetric extra U(1) models.
Without Froggatt-Nieslen mechanism, most stringent bound for proton decay channel
is given by $\tau(p\to e^+\pi^0)$. As the single quark interaction is doubly suppressed
by the VEV of gauge non-singlet flavon, g-quark life time become very long.
Therefore the allowed region  for flavon VEV is very narrow for TeV scale g-quark.

Introducing Froggatt-Nielsen mechanism, as we can weaken the $S_4$ flavon VEV suppression,
g-quark life time becomes short enough. 
From the naive power counting, we can expect
$p \to\mu^+K^0$ would dominate the proton decay width, however, corresponding operator
vanishes  by cancellation and do not contribute to proton decay.
Therefore our model predicts $p\to e^+K^0$ dominates the proton decay width.
This conclusion is not modified in Dirac neutrino model.

\appendix

\section{The detail of Eq.(34)}

The condition given in Eq.(34) gives four conditions  as follows
\eqn{
(\rho^2_4+\rho^2_3e^{2i\delta})e^{-i\phi}&=&
\rho^2_2e^{i\phi}-\frac{c^2_\nu-s^2_\nu}{c_\nu s_\nu}\rho_2\rho_4, \\
m_{\nu_1}e^{i\phi_1}&=&\rho^2_2e^{i\phi}-\rho_2\rho_4/t_\nu ,\\
m_{\nu_2}e^{i\phi_2}&=&\rho^2_2e^{i\phi}+\rho_2\rho_4t_\nu , \\
m_{\nu_3}&=&\rho^2_2.
}
These conditions gives seven equations for eleven unknown real variables
$\rho_{2,3,4},m_{\nu_{1,2,3}},\phi,\phi_{1,2},\delta,\theta_\nu$.
As there are three experimantal conditions for 
$m^2_{\nu_2}-m^2_{\nu_1},m^2_{\nu_2}-m^2_{\nu_3},\theta_\nu$,
therefore we must impose one external condition to fix all unknowns.
In this paper, we choose this external condition by hand as follow
\eqn{
\phi=0,
}
for simplicity. In this case the imaginary parts of Eq.(117)-(119)
become trivial.

%%%%%%%%%%%%%%%%%%%%%%%%%%%%%%%%%%%%%%%%%%%%%%%
%
%%%%%%%%%%%%%%%%%%%%%%%%%%%%%%%%%%%%%%%%%%%%%%%


\begin{thebibliography}{99}
\bibitem{SUSY}
H.~P.~Nilles, \PRP\vol{110}{1984}{1}.

\bibitem{extra-u1}D.~Suematsu and Y.~Yamagishi, \IJMP\vol{A10}{1995}{4521}.

\bibitem{s4u1}Y.~Daikoku and H.~Okada,
%{\em $S_4\times Z_2$ Flavor Symmetry in Supersymmetric Extra $U(1)$ Model}
 \PR\vol{D82}{2010}{033007}[arXiv:0910.3370[hep-ph]].

\bibitem{e6}F.~Zwirner, \IJMP\vol{A3}{1988}{49},
J.~L.~Hewett and T.~G.~Rizzo, \PRP\vol{183}{1989}{193}.

\bibitem{e6-FCNC}B.~A.~Campbell, J.~Ellis, K.~Enqvist, M.~K.~Gaillard and D.~V.~Nanopoulos,
\IJMP\vol{A2}{1987}{831};
%\cite{Daikoku:2010nj}\bibitem{Daikoku:2010nj}
  Y.~Daikoku and H.~Okada,
  %``Cancellation Mechanism of FCNCs in $S_4 x Z_2$ Flavor Symmetric Extra U(1) Model,''
  [arXiv:1008.0914 [hep-ph]].



\bibitem{s4pamela}Y.~Daikoku, H.~Okada and T.~Toma,
%{\em Two Component Dark Matters in $S_4\times Z_2$ Flavor Symmetric Extra $U(1)$ Model}
\PTP {\bf126} (2011) 855-883 [arXiv:1106.4717 hep-ph]].


\bibitem{OR}L.~O'Raifeartaigh, \NP\vol{B96}{1975}{331}, S.~P.~Martin [hep-ph/9709356].


\bibitem{R-sym}A.~E.~Nelson and N.~Seiberg, \NP\vol{B416}{1994}{46} [hep-ph/9309299].


\bibitem{domain-wall}F.~Riva, [arXiv:1004.1177[hep-ph]].


\bibitem{g-lifetime}M.~Kawasaki, K.~Kohri and T.~Moroi, \PR\vol{D71}{2005}{083502}
[astro-ph/0408426].


\bibitem{king}R.~Howl and S.~F.~King, JHEP\vol{0805}{2008}{008}[arXiv:0802.1909[hep-ph]].


\bibitem{kubo}J.Kubo,
%{\em Majorana phase in minimal $S_3$ invariant extension of the standard model}
\PL\vol{B578}{2004}{156}.
%156-164

\bibitem{t2k11} T2K Collaboration: K. Abe {\it et al.}, Phys. Rev. Lett.
{\bf 107}, 041801 (2011).
\bibitem{dc11} Double Chooz Collaboration: Y. Abe {\it et al.},
arXiv:1112.6353 [hep-ex].
\bibitem{daya12} Daya Bay Collaboration: F. P. An {\it et al.},
arXiv:1203.1669 [hep-ex].
\bibitem{reno12} RENO Collaboration: J. K. Ahn {\it et al.},
arXiv:1204.0626 [hep-ex].
%\cite{Tortola:2012te}
\bibitem{global-ana} 
  D.~V.~Forero, M.~Tortola and J.~W.~F.~Valle,
  %``Global status of neutrino oscillation parameters after recent reactor measurements,''
  arXiv:1205.4018 [hep-ph].
  %%CITATION = ARXIV:1205.4018;%%


\bibitem{PDG}Particle Data Group, \JP\vol{G37}{2010}{075021}
and 2011 partial update for the 2012 edition .


\bibitem{mass}Z-z.~Xing, H.~Zhang and Z.~Zhou, \PR\vol{D77}{2008}{113016} [arXiv:0712.1419[hep-ph]].



\bibitem{p-RGE}J.~Hisano,
%{\em Proton decay in the supersymmetric grand unified models} 
[hep-ph/0004266].

\bibitem{p-decay}P.~Nath, and P.~F.~Perez,
%{\em Proton Stability in Grand Unified Theories, in Strings and in Branes}
\PRP\vol{441}{2007}{191} [hep-ph/0601023].

\bibitem{PDform} T.~Goto and T.~Nihei, 
%{\em Effect of an RRRR dimension 5 operator on the proton decay in the minimal SU(5) SUGRA GUT model}
\PR\vol{D59}{1999}{115009}[hep-ph/9808255].

\bibitem{chiral}Y.~Aoki, C.~Dawson, J.~Noaki, and A.~Soni,
%{\em Proton decay matrix elements with domain-wall fermions}
\PR\vol{D75}{2007}{014507}[hep-lat/0607002].

\bibitem{fn}C.~D.~Froggatt and H.~B.~Nielsen, \NP\vol{B147}{1979}{277}.


\bibitem{newZp} ATLAS Collaboration, \PRL\vol{107}{2011}{272002} [arXiv:1108.1582[hep-ex]].


%\bibitem{g-quark}P.~Athron, J.~P.~Hall, S.~F.~King, S.~Moretti,
%D.~J.~Miller, R.~Nevzorov, S.~Pakvasa and M.~Sher,
%[arXiv: 1109.6373[hep-ph]].




%\bibitem{neutrino}T2K Collaboration: K.~Abe et.al, \PRL\vol{107}{2011}{041801}[arXiv:11062822[hep-ex]].
\end{thebibliography}
\end{document}